\documentclass[11pt,twoside]{article}
\usepackage{newpasp}

\usepackage{epsf}

\index{Weiner, B.J.}
\index{Vogel, S.N.}
\index{Williams, T.B.}

\def\HI{\hbox{H {\sc i}}}
\def\HII{\hbox{H {\sc ii}}}

\begin{document}

\title{Distance Constraints for High Velocity Clouds
from Optical Emission Lines}
\author{Benjamin J. Weiner}
\affil{Carnegie Observatories, 813 Santa Barbara St., Pasadena, CA 91101}
\author{Stuart N. Vogel}
\affil{Department of Astronomy, University of Maryland,
College Park, MD 20742}
\author{T.B. Williams}
\affil{Department of Physics \& Astronomy, Rutgers University,
136 Frelinghuysen Rd., Piscataway, NJ 08854}


\begin{abstract}
We report results from a survey of high velocity clouds 
and the Magellanic Stream for 
faint, diffuse optical recombination emission lines.
We detect H$\alpha$ emission with surface brightness
from 41 to 1680 mR from HVCs, and from $<40$ to 1360 mR
in the MS.  A simple model for the photoionizing 
radiation emergent from the Galaxy, normalized to the HVCs A and M with 
known distances, predicts distances from a few to 40 kpc, placing the
faintest HVCs in the Galactic halo, too far away for a Galactic fountain.
This model cannot explain the bright and spatially varying
H$\alpha$ in the Magellanic Stream, which requires another source of
ionization.  However, we do not find any HVCs
super-faint in H$\alpha$; even with another ionization source,
we conclude that the detected HVCs are not more than 2--4 times the 
distance of the MS (100-200 kpc).
\end{abstract}




\section{Introduction}

High velocity clouds are detected primarily in \HI;
no resolved optical counterparts such as stars have been
detected in HVCs, and the only distance upper limits
are for two HVCs seen in absorption against background halo stars.  
The lack of distance constraints makes the nature of 
and models for HVCs extremely uncertain; see the 
review of Wakker \& van Woerden (1997).  Currently favored
models include recycling of disk gas through a fountain
({\it e.g.}\ Bregman 1980); stripping from Galactic satellites;
and infall of possibly primordial 
gas ({\it e.g.}\ the Local Group model of Blitz {\it et al.}\ 1999).
These models place HVCs at from $< 10$ kpc to $\sim 1$ Mpc
respectively, a range of 100 in distance and $10^4$ in gas mass.

Faint, diffuse optical recombination emission lines are observed
from some HVCs.  H$\alpha$ emission must be caused by ionization on 
the cloud, either from photo-ionization by Lyman continuum
radiation, or another process such as collisional ionization.
Measurements of H$\alpha$ flux can constrain HVC distances,
given models for the ionization processes.

\section{Observations}

Observing diffuse emission lines from HVCs is difficult because
the emission is faint, night sky emission lines are strong and 
fluctuating, and HVCs are degrees across, larger than the field of view
of most optical telescopes.  Fabry-Perot observations of diffuse
emission provide large collecting solid angle with
moderate to high spectral resolution, needed to separate HVC emission
from sky lines.  FPs have previously been used to detect H$\alpha$ emission
from four HVCs and the Magellanic Stream (Weiner \& Williams 1996,
Tufte {\it et al.}\ 1998, Bland-Hawthorn {\it et al.}\ 1998).  
Chopping by several degrees between object and sky fields is 
necessary to obtain decent sky subtraction.

\begin{figure}
\plotfiddle{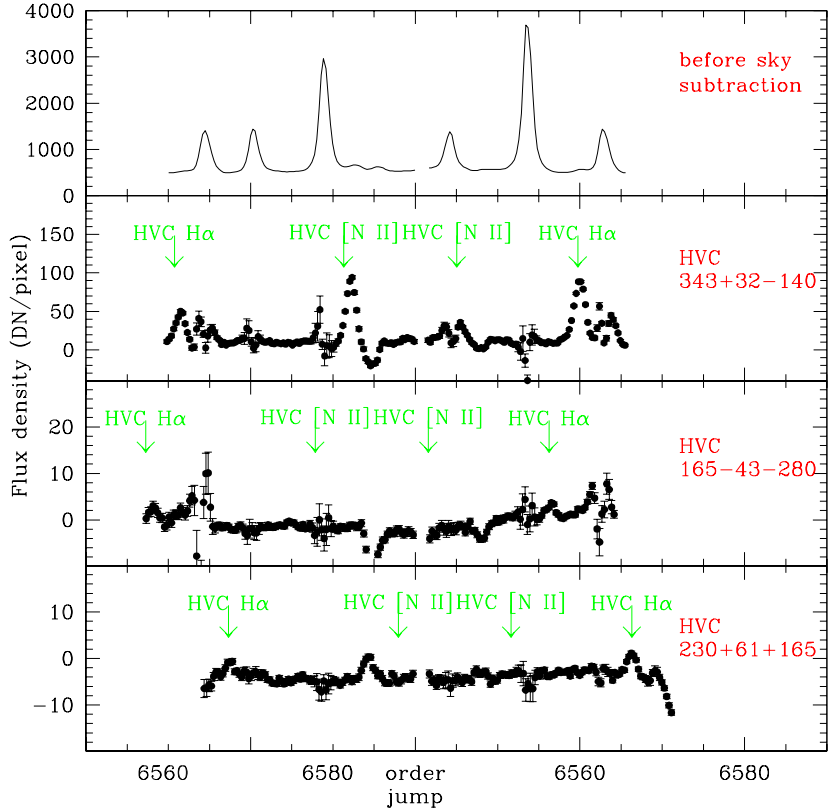}{4.0in}{0}{58}{58}{-176}{-98}
\caption{
Las Campanas Fabry-Perot spectra of HVCs.  Note that the wavelength 
scale has a break (an order jump).  The raw spectrum (top panel)
is dominated by night sky lines; residuals from these
and Galactic [N~{\sc ii}] are seen in the sky-subtracted spectra.
Arrows indicate the optical wavelengths
expected given the radio \HI\ HVC velocity.
}
\end{figure}

Figure 1 shows example spectra from the Wide Field Camera F-P
at the Las Campanas 2.5-m, with a field of view of 25$^\prime$
and etalon FWHM 1.2 \AA\ ($R=5500$).  The top panel is a
extracted spectrum of a 900 sec exposure of HVC 343+32--140, one of the 
HVCs brightest in H$\alpha$, before subtraction of a sky-field spectrum.
The second panel is after a two-step sky subtraction: sky lines
in the object-field and sky-field spectra are fit and subtracted, then 
the sky-field continuum is subtracted from the object-field continuum.
The H$\alpha$ flux is strong, 1060 milli-Rayleighs
(mR), as is [N~{\sc ii}] 6583, with [N~{\sc ii}]/H$\alpha$ = 1.1.
(1 Rayleigh = $10^6$ photons cm$^{-2}$ s$^{-1}$ into $4\pi$.)
The lower two panels show two HVCs very faint in H$\alpha$, 
HVC 165--43--280 (41 mR) and HVC 230+61+165 (48 mR).  
The H$\alpha$ detections agree well with the \HI\ velocities.  
No [N~{\sc ii}] is detected.  Note the tremendous difference in 
strength of HVC H$\alpha$ and night-sky lines.

\begin{figure}
\plotfiddle{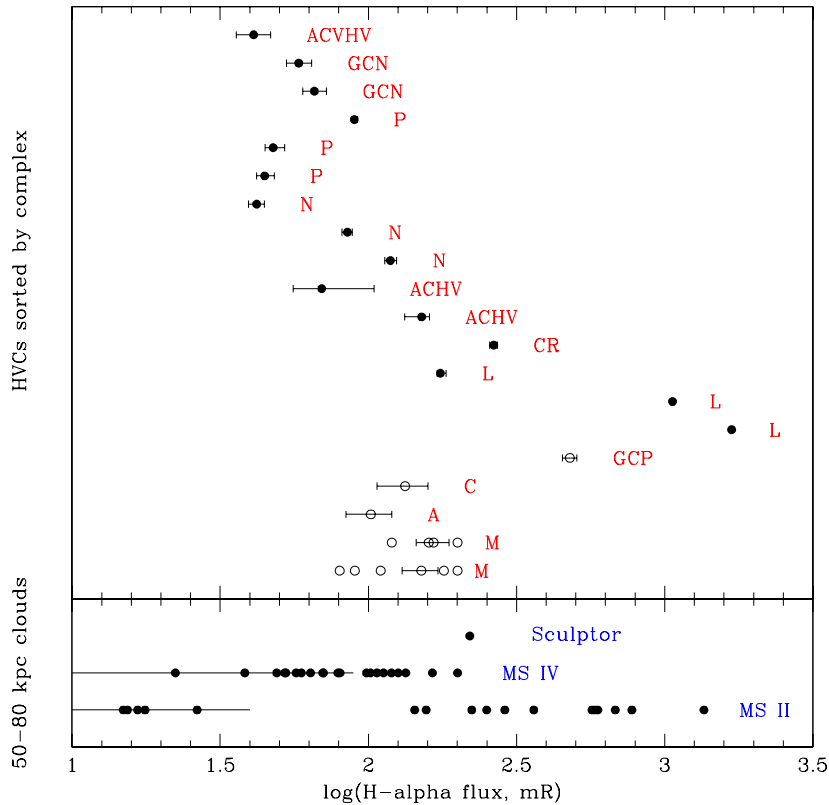}{4.0in}{0}{58}{58}{-176}{-98}
\caption{
H$\alpha$ emission from HVCs (upper panel), and the Magellanic Stream
and Sculptor DSph (lower panel).
The HVCs are grouped by complex (ACVHV, GCN, etc.; Wakker \& van Woerden 1991)
along the y-axis, each line being one HVC.  Open circles:
M, A, and C from Tufte {\it et al.}\ (1998),
and GCP from Bland-Hawthorn {\it et al.}\ (1998).
Filled circles: new data from LCO and CTIO.
}
\end{figure}

Figure 2 compiles our results from the LCO FP and the Rutgers FP at
the CTIO 1.5-m, and HVC H$\alpha$ detections from Tufte {\it et al.}\
(1998) and Bland-Hawthorn {\it et al.}\ (1998).  There is a wide
range of HVC H$\alpha$ intensity, but clouds in the same complex tend
to have similar intensities, which suggests that the variations
between complexes are due to HVC properties ({\it e.g.}\ distances)
rather than extrinsic variations ({\it e.g.}\ spatial variations in
the ionizing field escaping from the Galaxy).  On the other hand, the
Magellanic Stream points vary widely: some points have strong emission
while others have weak or no H$\alpha$ despite a high \HI\ column
density.  Strong emission in the MS is often located near cloud edges.

A fundamental result of our H$\alpha$ survey is that we have not found
any cloud near or below our photon-limited noise (generally
$10-20$ mR at 2$\sigma$).  The faintest
H$\alpha$ detections, at 41--50 mR, are well above our noise limit.
We always either detect H$\alpha$ (15 of 20 HVCs), or are
clobbered by residuals from sky-subtraction when HVCs are not well 
separated from sky lines, raising the detection limit.

\section{Models for the source of the emission}

If H$\alpha$ emission from HVCs is due to photo-ionization by flux
from the Galaxy, we can infer distances to the HVCs.  We use the HVCs
A and M with known distance brackets (4--10 kpc and $<5$ kpc; van
Woerden {\it et al.}\ 1999, Danly {\it et al.}\ 1993) and H$\alpha$
fluxes (Tufte {\it et al.}\ 1998) to normalize a model for the
ionizing flux escaping from the Galaxy.  Figure 3 shows contours of
the ionizing photon flux $F_{LC}$ in the model; it has a total
ionizing luminosity of $L_{LC} = 2.7 \times 10^{53}$ photons s$^{-1}$,
distributed in an exponential disk, and
models the Galactic absorbing layer with 
a one-sided face-on
optical depth to ionizing photons of $\tau = 2.5$, yielding an
overall, angle-averaged escape fraction $<f_{esc}>\ = 2\%$.    
(See also Bland-Hawthorn
\& Maloney 1999 and Bland-Hawthorn, these proceedings.)

The inferred distances of HVCs
are indicated; the error bars are for systematic variations by
a factor 1.5 up or down in H$\alpha / LC$ ratio (statistical errors on the
fluxes are much smaller).  The brightest clouds are within a few kpc
of the Galactic plane but fainter clouds are well away from the plane.
These clouds at $|z|>10$ kpc are inconsistent with a Galactic fountain
origin, especially given their high velocities.

However, the observed H$\alpha$ in the Magellanic Stream cannot be 
explained by this model; MS II and MS IV
are much too bright compared to HVCs A and M. 
In fact, at a distance of $D \sim 50$ kpc, the MS H$\alpha$ emission
cannot be explained by a reasonable model of photoionization
from the Galaxy.  The H$\alpha$ photon flux is 
$F(H\alpha) = 0.46~F_{LC} 
\simeq 0.46~L_{LC}/(4\pi D^2)~e^{-\tau\,{\rm csc}\,b}$, 
so that H$\alpha$ of 300--600 mR at MS II requires $\tau = 0.3-0$ and
$<f_{esc}>\ = 60 - 100\%$.  The required escape fraction is
unrealistically large (even if $L_{LC}$ were increased by $\times 2$) 
since most ionizing radiation must be absorbed in the Galaxy to power \HII\
regions.  (In agreement with the authors, we find that Bland-Hawthorn
\& Maloney 1999 overestimated the ionizing flux incident on the MS.)
Furthermore, there are strong spatial variations in the MS
H$\alpha$, shown by the large dispersion of points in Figure 2.
Photoionization from the Galaxy should produce roughly the same
H$\alpha$ flux on any \HI\ surface optically thick to ionizing radiation,
since the ionizing photons travel nearly along our line of sight to the MS.
Factors of 2--3 variation between different points are tolerable
either from variations in the incident field or geometry of the
absorbing surface, but factors of 30 are not.

A second source of ionization in the Galactic halo is needed to
explain the MS H$\alpha$. Simply put, why are some points in the Magellanic
Stream brighter than Complexes A and M, despite being much further away?
The only likely source is collisional ionization, probably from
interaction with hot halo gas through ram pressure and
turbulent mixing/thermal conduction (Weiner \& Williams 1996).
The MS radial velocities are large and the space velocities 
with respect to halo gas are probably above 200 km~s$^{-1}$;
energy input from ram pressure goes as $v^3$, so it may
explain the brightness of the MS with respect to HVCs A and M,
but it is barely adequate for reasonable halo gas densities
(Weiner \& Williams 1996); perhaps the Magellanic Stream
clouds are colliding with each other.
Collisional ionization is required, but the brightest
MS points remain a puzzle.

Recent FUSE detections of O~{\sc vi} absorption in the Magellanic Stream and
in a few HVCs provide further evidence for collisional ionization in
HVCs (Sembach {\it et al.}\ 2000).  O~{\sc vi} can be produced
collisionally in hot gas at $T \sim 10^{5.5}$ K, which could arise
from the interaction of an \HI\ cloud with hot halo gas, but O~{\sc vi}
cannot realistically be produced in HVCs by photoionization from O
stars.

\begin{figure}
\plotfiddle{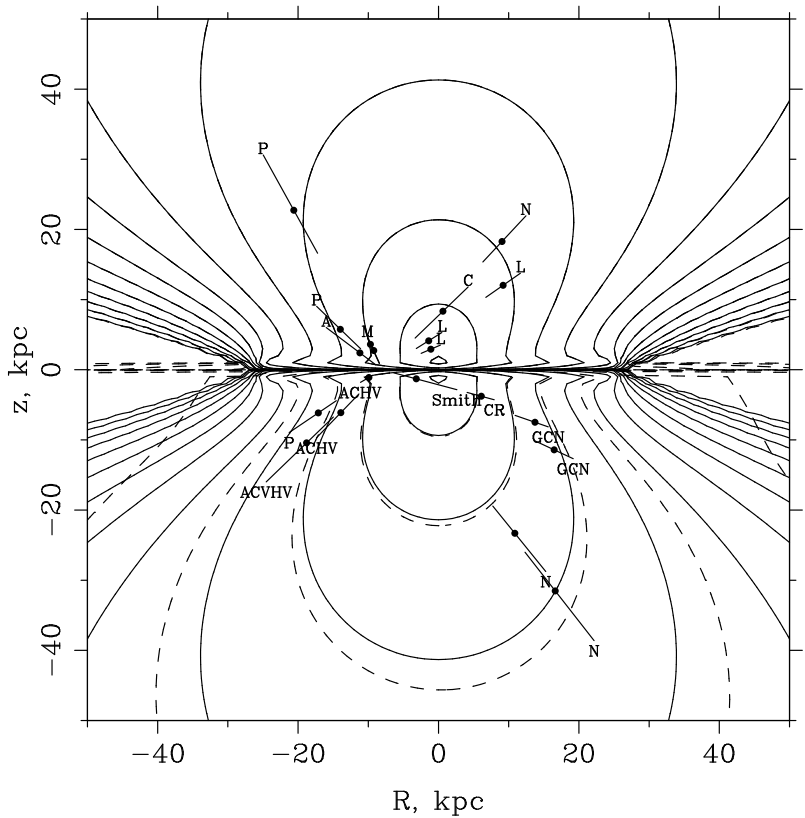}{3.85in}{-90}{62}{62}{-242}{348}
\caption{
A simple model of the ionizing flux emergent from the Galaxy,
with contours of $F_{LC}$ from 1 to $10^{6.5}$ photons cm$^{-2}$ s$^{-1}$ 
by 0.5 dex (dashed contours include the LMC).
Positions of HVCs from the simple photo-ionization model are indicated
(rotated onto the $\ell= 180\deg - 0\deg$ plane, so that $R$ is
represented accurately).}
\end{figure}

\section{Conclusions}

Given the existence of a second source of ionization in the Galactic
halo, simple photoionization models may not yield reliable
distances.  However, robust predictions can still be made for
a relation of H$\alpha$ flux to distance.

HVCs which are bright in H$\alpha$ (as HVCs go)
are likely only a few kpc from the Galactic plane.
The brightest HVCs plotted in Figure 2 (L, GCP, CR) have
inferred distances which give them 
low ``deviation velocities'' (Wakker \& van Woerden 1991), so they are
not too far from Galactic differential rotation, 
despite having high $V_{LSR}$; these HVCs could well be recycled disk gas.
High [N~{\sc ii}]/H$\alpha$ ratios suggest high-metallicity gas
photoionized by a hardened radiation field, as in the extraplanar
diffuse ionized gas of NGC 891 (Rand 1998).

Conversely, if an HVC is very far from the Galaxy, the only source 
of ionization is the metagalactic UV background.  Then the HVC should
be very faint in H$\alpha$, approaching the upper limits on
H$\alpha$ emission from isolated extragalactic H~I clouds:
$<20$ mR at $2\sigma$ from the Giovanelli-Haynes cloud
(Vogel {\it et al.}\ 1995) and $<30$ mR at $2\sigma$ from the Leo Ring
(Donahue {\it et al.}\ 1995).

We do not find any HVCs that are very faint, 
even though our photon-limited noise is generally 10--20 mR (2$\sigma$): 
we either detect H$\alpha$, or are limited by
sky subtraction residuals.   The faintest HVCs we find are
165--43--280, 47--52--129, 227--34+114, and 230+61+165, with H$\alpha$
emission at 41--48 mR.  
In the photoionization model of Figure 3 their distances are 
$D= 15, 40, 11$, and 26 kpc from the Sun.  We can push the 
distances to the limit by considering solely collisional ionization:
the HVC H$\alpha$ fluxes are some 5-15 times smaller
than the fluxes typical of the Magellanic Stream.  If we attribute
the H$\alpha$ to interaction with hot halo gas, and assume that the 
hot gas density falls off as $r^{-2}$ to $r^{-3}$, the HVCs can be at
most 2 to 4 times farther than the MS, or 100-200 kpc.

At distances of somewhere between 20 and at most 200 kpc, the
H$\alpha$-faint HVCs are not Local Group objects as defined by 
the Blitz {\it et al.}\ (1999) model.  However, 
their low [N~{\sc ii}]/H$\alpha$ ratios suggest low metallicity,
and they are not
easily produced by a Galactic fountain: for example,
HVC 165--43--280, with $V_{GSR} = -240$ km~s$^{-1}$,
has faint H$\alpha$, is at least 10 kpc away (20 kpc Galactocentric
radius), and has a large
mass and kinetic energy; perhaps it is infalling tidal debris.
The best explanation for these H$\alpha$-faint HVCs is still likely to be
gas, of uncertain origin, accreting onto the Galaxy.




\end{document}